\documentstyle[11pt,newpasp,twoside,epsf]{article}
\def\edcomment#1{\iffalse\marginpar{\raggedright\sl#1\/}\else\relax\fi}
\reversemarginpar

\begin{document}
\title{AGN Populations from Optical Identification of {\it ASCA}
Surveys}
\author{Masayuki Akiyama}
\affil{Subaru Telescope, National Astronomical Observatory of Japan,
Hilo, HI, 96720}
\author{Yoshihiro Ueda}
\affil{ISAS, Sagamihara, Kanagawa, 229-8510, Japan}
\author{Kouji Ohta}
\affil{Department of Astronomy, Kyoto University, Kyoto, 606-8502, Japan}

\begin{abstract}
To understand luminous AGNs in the
$z<1$ universe, the {\it ASCA} AGN samples are the best at present.
Combining the identified sample of AGNs from {\it ASCA} Large
Sky Survey and Medium Sensitivity Survey, the sample of hard X-ray
selected AGNs have been expanded up to 108 AGNs above the flux
limit of 10$^{-13}$ erg s$^{-1}$ cm$^{-2}$ in the 2--10~keV hard
X-ray band. We discuss the fraction of absorbed AGNs in the hard 
X-ray selected AGN sample, and nature of absorbed luminous AGNs.
\end{abstract}

\section{Introduction : Importance of a Bright Hard X-ray AGN sample}

\begin{figure}
\plotone{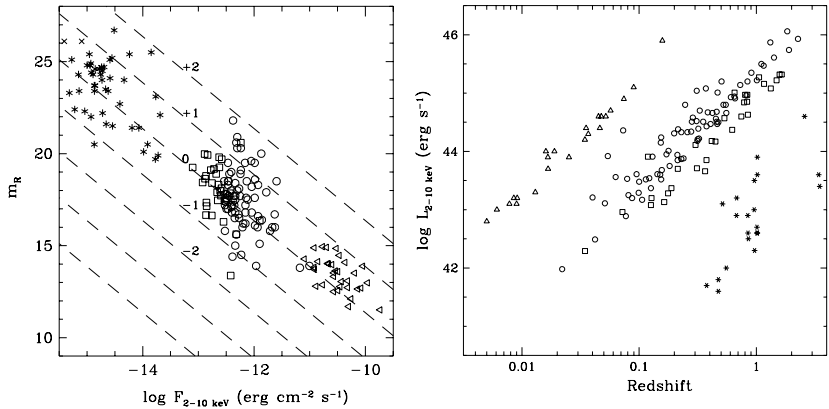}
\caption{
Left) $R$-band magnitudes of optical counterparts of ALSS (square)
and AMSSn (circle) AGNs are plotted as a function of 2--10~keV
hard X-ray flux. Dashed lines represent the X-ray to optical flux
ratio of $\log f_{X}/f_{V} = +2$,$+1$,0,$-1$,and $-2$ from top to bottom.
Triangles and
asterisks indicate samples from {\it HEAO1} A2 (Piccinotti et al. 1982) 
and {\it Chandra} survey in HDF-N (Hornschemeier et al. 2001)
Right) Hard X-ray luminosities of the hard X-ray selected
AGNs plotted as a function of redshift. Same symbols as in the left
panel.}
\end{figure}

The fraction of absorbed AGNs, especially luminous absorbed
AGNs, is one of a
big issue in understanding the true number density of active nuclei
in the universe. Recently many candidates of absorbed luminous AGNs
are found in AGN surveys in radio, X-ray, and near-infrared wavelengths
(e.g., Webster et al. 1995).
The discoveries imply that
we have been missing significant fraction of nucleus with high activity
in traditional optical/UV-selections of AGNs due to absorption to
the nucleus. However the fraction of the absorbed AGN in the entire
AGN population is not clear. Radio-selected samples are affected by
red AGNs with red synchrotron component (Francis et al. 2001), 
soft X-ray selection is biased
against heavily absorbed AGNs (Kim \& Elvis 1999), and 
2MASS-selected red AGNs are limited in the low redshift 
universe (Cutri et al. in this volume).

In order to construct a complete sample of AGNs less biased against
absorption to nucleus, we conduct optical follow-up observations
for {\it ASCA} Large Sky Survey (hereafter ALSS; Ueda et al. 1999)
and {\it ASCA} Medium Sensitivity Survey 
(hereafter AMSS; Ueda et al. 2001) in the hard X-ray band.
Hard X-ray emission can penetrate the obscuring matter of absorbed AGNs
and is very suitable in searching absorbed AGNs. 
Using 2--10~keV hard X-ray
emission, we can detect AGNs with X-ray absorption up to hydrogen column
density of 10$^{22\sim23}$ cm$^{-2}$, which corresponds to $A_{V}$ of 
20 $\sim$ 50 with galactic conversion factor, without bias. 
ALSS is a survey in a continuous field with 5.4 square degree
near the north galactic pole. We selected 34 X-ray sources detected
with SIS 2--7~keV significance larger than 3.5$\sigma$. The sources
are identified with 30 AGNs, 2 clusters of galaxies and 1 galactic star
(Akiyama et al. 2000). One X-ray source with hard spectrum is still 
unidentified, and {\it Chandra} follow-up observation is planed in 
Cycle 3.
AMSS is a serendipitous source survey based on 
{\it ASCA} pointing observations conducted in 
high galactic latitude region ($|b|>20^{\circ}$).
We conducted optical follow-up observations for 86 X-ray sources
detected with GIS 2--10~keV significance larger than 5.6$\sigma$
in the northern sky (declination above $20^{\circ}$; 
we call AMSSn sample). All of the
X-ray sources are identified with 78 AGNs, 7 clusters of galaxies, 
and 1 galactic star (Akiyama et al. in preparation). 
In total, we constructed sample of 108 hard X-ray selected AGNs 
with the flux limit of {\it ASCA},
about $\sim 10^{-13}$ erg s$^{-1}$ cm$^{-2}$ in the 2--10~keV band.
In Figure 1, we plotted the hard X-ray flux vs. optical magnitude (left)
and the redshift vs. luminosity distribution (right) diagrams of
ALSS and AMSSn AGNs. The {\it ASCA} samples are 
two orders of magnitude brighter and more luminous than the sample of 
deep {\it Chandra} and {\it XMM-Newton} surveys, and
consists of luminous AGNs, i.e., QSOs,  in the universe below redshift 1.
The high completeness of the {\it ASCA} samples makes us possible
to discuss the fraction of absorbed AGNs definitely.

\section{Fraction of Heavily Absorbed AGNs}

Using the hardness of X-ray spectrum of each source, 
we can estimate the X-ray absorption to the nucleus 
in each object. The 0.7--10~keV apparent photon index distributions 
of ALSS and AMSSn AGNs are plotted as a function of redshift in left 
panel of Figure 2. The upper and lower solid lines in the figure 
correspond to the apparent photon index of an object 
with intrinsic photon index of 1.7 and X-ray absorption with hydrogen
column density of $\log N_{\rm H}=22$(cm$^{-2}$) and $\log N_{\rm H}=23$(cm$^{-2}$) at each redshift,
respectively. The X-ray sources with apparent photon index smaller
than 1 can be regarded as significantly harder than canonical power-law
spectra of broad-line AGNs (with photon index of 1.7). They 
correspond to intermediate redshift
AGNs with X-ray absorption of $\log N_{\rm H}=22-23$(cm$^{-2}$) and
high-redshift AGNs with absorption of $\log N_{\rm H}>23$(cm$^{-2}$).
At high-redshift ($z\sim1$), 
the apparent photon indexes
of highly absorbed objects become close to that of object without 
absorption,
because we observe very high energy photons of the source-frame, which
are less affected by absorption than low-energy photons.

\begin{figure}
\plottwo{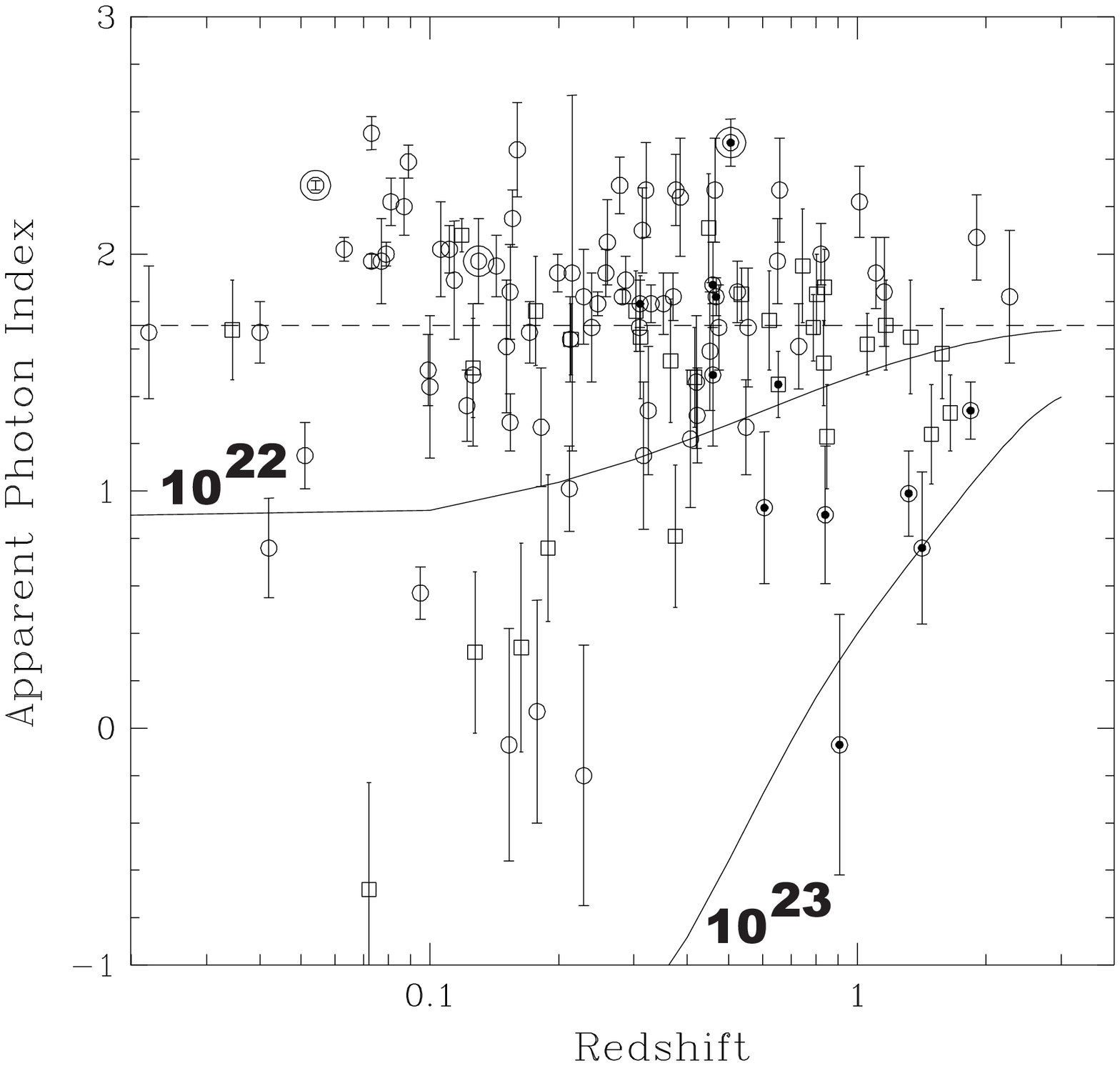}{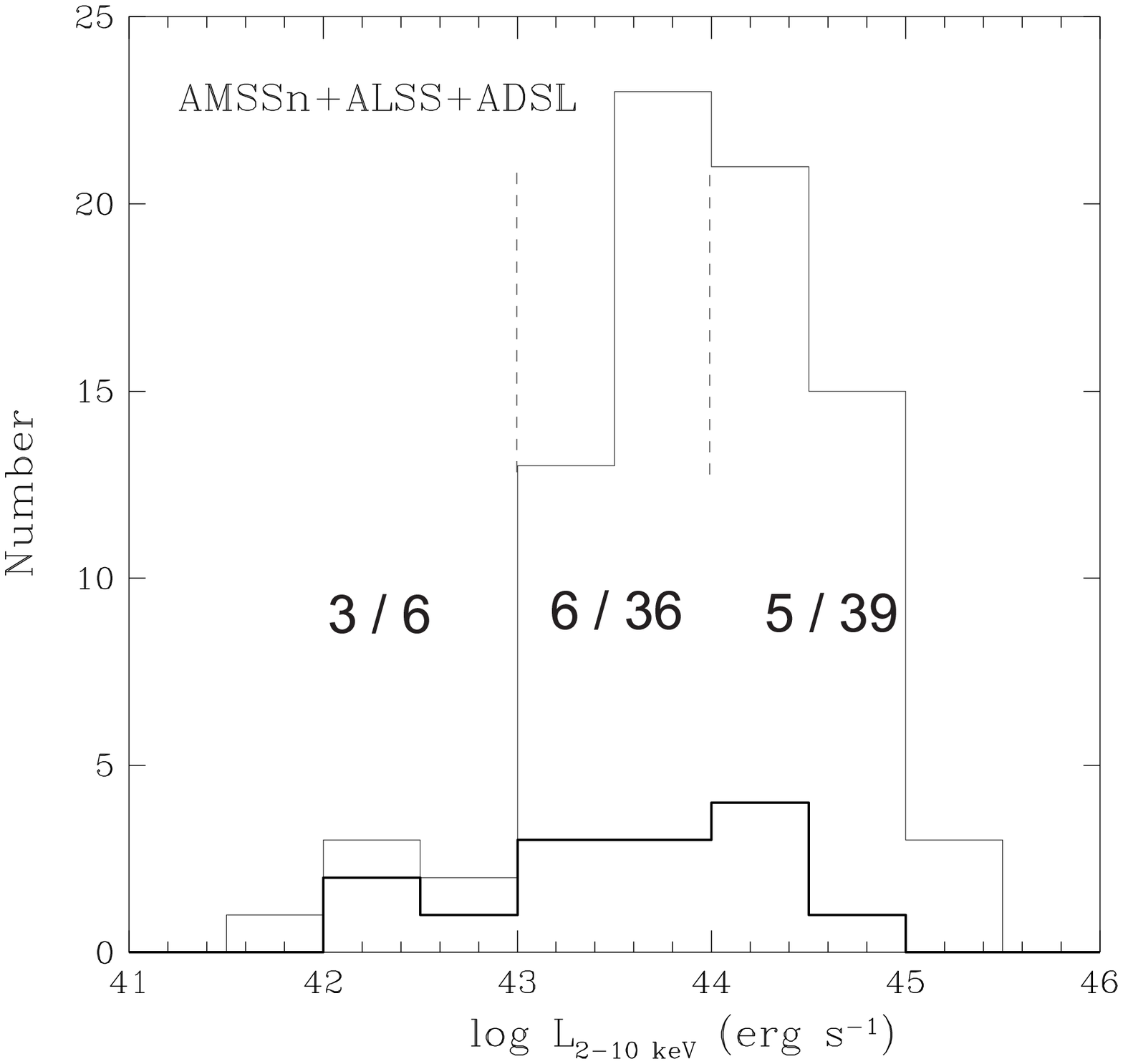}
\caption{
Left)
Apparent photon index of ALSS (squares) and AMSSn (circles) AGNs in the
0.7--10~keV hard X-ray band are plotted as a function of
redshift. BL Lac objects are marked with large circles. 
The solid lines show the apparent photon index
of power-law continuum with intrinsic photon index of 1.7
absorbed by hydrogen column density of $10^{22}$ cm$^{-2}$ (top) and
$10^{23}$ cm$^{-2}$ (bottom) at each redshift. 
AGNs with a faint optical counterpart
($\log f_{X}/f_{V}$ larger than 1) are marked with dots.
Right) 
Luminosity distribution of all (thin) and
significantly absorbed (thick) AGNs below redshift of 0.6
from combination of AGN samples of ALSS, AMSSn, and ADSL.
}
\end{figure}

Based on the estimated amount of absorption to nucleus, we
examine the fraction of absorbed AGNs in the hard X-ray selected AGNs.
For simplicity, we limit the sample below
redshift of 0.6, and regard AGNs with 
$\log N_{\rm H}>22$(cm$^{-2}$) as significantly absorbed AGNs.
It should be noted that at high-redshifts ($z>0.6$) AGNs with hydrogen
column density of $\log N_{\rm H} = 10^{22-23}$ (cm$^{-2}$)
can not be regarded as significantly absorbed in the current sample.
In right panel of 
Figure 2, the luminosity distribution of the all AGNs and 
significantly absorbed AGNs from combination of the ALSS, AMSSn, and
{\it ASCA} Deep Survey in the Lockman Hole
(ADSL; Ishisaki et al. 2001) is plotted.
The fraction of absorbed AGNs is higher in the lowest luminosity
range, but there is no clear deficiency of absorbed AGN 
above 10$^{44}$ erg s$^{-1}$, which is observed in the ALSS sample
(Akiyama et al. 2000).
The fraction of absorbed AGN are 6/36 and 5/39 in the
luminosity between 10$^{43}$ erg s$^{-1}$ and 10$^{44}$ erg s$^{-1}$
and in the luminosity above 10$^{44}$ erg s$^{-1}$, respectively.
The fraction of absorbed AGNs is higher in the luminosity range
below 10$^{43}$ erg s$^{-1}$ (3/6) than in the luminosity range
above, but the number of AGNs in the low luminosity range is 
fairly limited. 
The fraction of luminous ($L_{\rm X} > 10^{44}$ erg s$^{-1}$)
AGNs with $\log N_{\rm H} > 22$(cm$^{-2}$)
in the sample of AGNs without bias up to 
$\log N_{\rm H} = 23$(cm$^{-2}$) ($\sim$15\%) is 
clearly smaller than that expected
from the models of cosmic X-ray background 
(45\%; Comastri et al. 1995) or that observed in 
local low-luminosity Seyfert galaxies (40\%; Risaliti et al. 1999). 

\section{Case studies on absorbed QSOs}

The fraction of absorbed QSOs is not as large as expected,
but we detected several candidates of absorbed QSOs in {\it ASCA}
surveys. Their counterparts are relatively faint and have larger
X-ray to optical flux ratio than normal AGNs (see dotted objects
in left panel of Figure 2). Most of high-redshift AGNs with hard
X-ray spectra have large X-ray to optical flux ratios.
The X-ray to optical flux ratio is
similar to those of optically-faint hard X-ray source population 
found in deep {\it Chandra} surveys (see left panel of Figure 1;
e.g., Alexander et al. 2001),
and the {\it ASCA} optically-faint AGNs can be low-redshift and/or
high-luminosity cousin of the {\it Chandra} population. 

Although measured amount of X-ray absorption is large,
most of the luminous absorbed QSOs show broad MgII 2800{\AA} or
H$\alpha$ 6563{\AA}
emission line. The origin of the discrepancy can be
1) broad MgII 2800{\AA} from scattered nuclear light
or 2) discrepancy between amount of X-ray photoelectric
absorption and optical dust reddening. We show two examples
of absorbed QSOs fall in each category.

\subsection{An absorbed QSO at $z=0.65$ with a strong 
broad MgII 2800{\AA} emission line}

\begin{figure}
\plotone{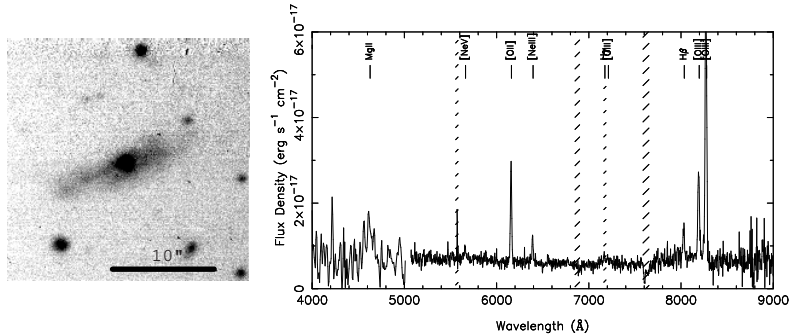}
\caption{Left) Optical $R$-band image of AX~J131831+3341.
Right) Optical spectrum of AX~J131831+3341. A 1800s 
FOCAS/Subaru spectrum above 5000{\AA} with 
3600s MOSCA/Calar Alto 3.5m spectrum
around MgII 2800{\AA} emission.}
\end{figure}

AX~J131831+3341 is an absorbed radio-quiet QSO at
a redshift of 0.65 found in ALSS (Akiyama et al. 2000). 
Its X-ray luminosity is estimated to be $\sim 10^{45}$ erg s$^{-1}$,
which corresponds to the luminosities of QSOs. 
The observed X-ray spectrum of the object in
a 0.7--10~keV band is described by intrinsic absorption with a
hydrogen column density of $N_{\rm H}=6.0^{+4.4}_{-4.2} \times
10^{21}$ cm$^{-2}$ and an intrinsic photon index of 1.7. 
The hydrogen column density 
corresponds to the lower edge of the column density distribution of
Seyfert 1.8-1.9 galaxies.

The optical spectrum of the object shows strong emission 
lines, such as broad MgII 2800{\AA},
narrow [OII] 3727{\AA}, and narrow [OIII] 5007{\AA}, but
no broad H$\beta$ emission line (see right panel of Figure 3). 
Its small H$\beta$-to-[OIII] 5007{\AA}
equivalent width ratio is comparable to those of Seyfert 1.8-2 galaxies.
Optical and near-infrared images show nuclear and extended
components (see left panel of Figure 3). Because the nuclear component
has very red $I-K$ color but blue $V-R$ and $R-I$ colors,
the nucleus is thought to be 
absorbed with $A_V\sim3$ and emerge only in the $K$-band
(Akiyama and Ohta 2001). The amount of absorption is consistent
with the amount of X-ray absorption. 
The optical blue continuum and broad MgII 2800{\AA} emission line
can originate from scattered nuclear light (Akiyama et al. 2001).

\subsection{A candidate of a type-2 QSO with 
large X-ray absorption and a strong broad-H$\alpha$ emission line}

AX~J08494+4454 is a candidate of a type-2 QSO at $z=0.9$ 
found in the course of the optical identification of {\it ASCA}
deep survey in Lynx field (Ohta et al. 1996). 
Recently, deep {\it Chandra} hard X-ray spectrum and IRCS/Subaru
$J$-band spectrum of the object are obtained (Akiyama et al. 2002).
The 0.5--10~keV 150ks {\it Chandra} spectrum of AX~J08494+4454
is hard, and is explained well with a power-law continuum absorbed
by a hydrogen column density of $(2.3\pm1.1) \times 10^{23}$ cm$^{-2}$.
The 2--10~keV luminosity of the object is estimated to be
$7.2^{+3.6}_{-2.0} \times 10^{44}$ erg s$^{-1}$, after correcting
the absorption, and reaches hard X-ray luminosities of QSOs.
The large X-ray absorption  and the
large intrinsic luminosity support the original identification of
AX~J08494+4454 as a type-2 radio-quiet QSO.
Nevertheless, deep Subaru/IRCS $J$-band spectroscopic
observation suggests the presence of a strong 
broad H$\alpha$ emission line from AX~J08494+4454
(left panel of Figure 4). 
The broad H$\alpha$ emission line has a velocity width of
$9400\pm1000$ km s$^{-1}$, which corresponds to a typical
broad-Balmer line velocity width of a luminous QSO.
The existence of the strong broad H$\alpha$ line means that
the object is not a type-2 QSO, but a luminous cousin of
a Seyfert 1.9 galaxy in the source-frame optical spectrum.
The Balmer decrement of broad lines,
the broad H$\alpha$ emission to the hard X-ray luminosity ratio,
and optical SED (right panel of Figure 4)
suggest that the nucleus is affected by dust extinction with
$A_V$ of $1-3$ mag in the optical wavelength.
The estimated amount of dust extinction is much smaller than that
expected from the X-ray column density ($A_V=130\pm60$ mag).
The discrepancy can be explained with a smaller dust to gas
mass ratio which may due to dust sublimation in the X-ray absorbing
matter, the size difference between optical and X-ray emitting
region, or different dust size distribution in AGNs 
(e.g., Maiolino et al. 2001).

\begin{figure}
\plotone{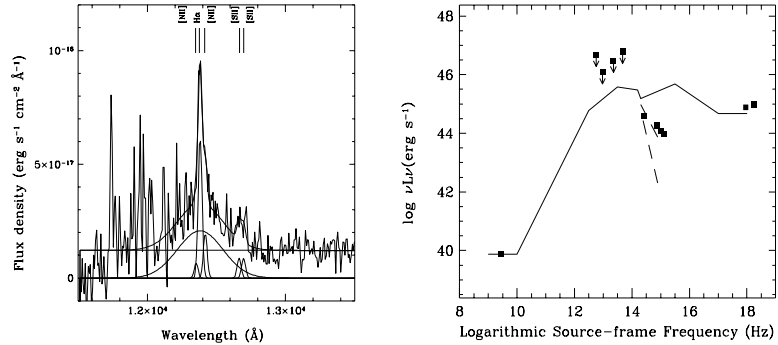}
\caption{
Left) $J$-band spectrum of AX~J08494+4454. 
The best fit models are also plotted with solid lines.
Right) Spectral energy distribution (SED) of AX~J08494+4454.
Solid line indicates SED of an average radio-quiet QSO (Elvis et al. 1994)
and is normalized at the data point observed at 1.4 GHz.
Dashed lines represent optical SEDs affected by dust extinction 
with $A_V$ of 1 mag (upper) and 2 mag (lower).
}
\end{figure}

\acknowledgements

The authors would like to thank ALSS and AMSS members.

\end{document}